# Personalized Adaptive Cruise Control and Impacts on Mixed Traffic

Mehmet Fatih Ozkan and Yao Ma

*Abstract*— **This paper presents a personalized adaptive cruise control (PACC) design that can learn human driver behavior and adaptively control the semi-autonomous vehicle (SAV) in the car-following scenario, and investigates its impacts on mixed traffic. In mixed traffic where the SAV and human-driven vehicles share the road, the SAV's driver can choose a PACC tuning that better fits the driver's preferred driving strategies. The individual driver's preferences are learned through the inverse reinforcement learning (IRL) approach by recovering a unique cost function from the driver's demonstrated driving data that best explains the observed driving style. The proposed PACC design plans the motion of the SAV by minimizing the learned unique cost function considering the short preview information of the preceding human-driven vehicle. The results reveal that the learned driver model can identify and replicate the personalized driving behaviors accurately and consistently when following the preceding vehicle in a variety of traffic conditions. Furthermore, we investigated the impacts of the PACC with different drivers on mixed traffic by considering time headway, gap distance, and fuel economy assessments. A statistical investigation shows that the impacts of the PACC on mixed traffic vary among tested drivers due to their intrinsic driving preferences.**

## I. Introduction

Over the past decade, the growing vehicle technology in the transportation sector has enabled the development of Advanced Driver Assistance Systems (ADAS) [1]. Adaptive Cruise Control (ACC) is one of the essential features of the ADAS that helps drivers to reduce their burden and to improve driving safety during the traffic operations by controlling the longitudinal motion of the vehicle [2-3]. The ACC implementations have demonstrated an efficient control strategy to boost driving comfort and safety for different longitudinal driving operations, such as maintaining a fixed distance and a user-defined speed setpoint [4]. While the current ACC technology is at an advanced level, the development of the ACC adaptation to the drivers' preferences and driving skills is an area of research with considerable importance [5].

Existing studies use different mathematical models that mimic a human driver's driving behaviors to provide personalized ACC (PACC) in the dynamic traffic scenarios [6-9]. In [8], the authors propose a multimodel PACC approach to learn drivers' preferred driving behaviors such as highway driving preferences by extracting the numerical driving style indicators from driving data and to execute driver-specific control actions by using the model predictive control (MPC) design. The results suggest that the proposed MPC design can accurately learn and replicate the personalized driving behaviors while guaranteeing the safe operation of the vehicle during the traffic conditions.

To capture the drivers' preferred driving behaviors, it is necessary to establish an accurate and generalizable driver behavior model to achieve an efficient PACC design. Some studies suggest that driver behaviors could be effectively learned through Inverse Reinforcement Learning (IRL) approach [10-12] using the driver's demonstrated driving data. The objective of the IRL approach is to replicate the observed driving style best with a cost function, which consists of a linear combination of features such as preferred speed and gap distance, and their corresponding weights. In [10], the authors use the IRL approach to learn individual driving styles with a cost function that captures the unique highway driving properties of the driver and efficiently computes driver-specific trajectories for the vehicle in autonomous mode. The experimental results suggest that the learned model is suitable to autonomously control the vehicle with similar observed driving characteristics from human drivers in different traffic conditions.

Although it is shown that PACC strategies can learn human driver behavior and control the vehicle with similar observed driving behaviors of human drivers, the impacts of such PACC design on mixed traffic have not been comprehensively assessed yet. Since manual driving vehicles will remain to be primary vehicles as the popularity of semi-autonomous vehicles (SAVs) is a gradual development process, a mixed traffic scenario where SAVs will co-exist with manual driving vehicles is expected to be a common norm for the ground transportation in the long period of time. Therefore, insights into the impacts of SAV with PACC design on mixed traffic will be valuable for practical and theoretical significance before large-scale deployment of the PACC systems in the ground transportation sector.

The objective of this study is to design a personalized ACC system that can learn and mimic the driving style of the human driver in the car-following scenario, and to investigate its impacts on mixed traffic considering the diverse characteristics of the human driver's behaviors.

The distinct contributions of this study include: 1) a driver behavior learning model based on inverse reinforcement learning is designed to capture the driving preferences of the different human drivers from the demonstration. 2) a personalized ACC design for the semi-autonomous vehicle is proposed to execute the best driver model to replicate an individual's driving style in a variety of realistic driving scenarios. 3) the impacts of such a personalized ACC

M. F. Ozkan and Y. Ma are with the Department of Mechanical Engineering, Texas Tech University (e-mail: mehmet.ozkan@ttu.edu and yao.ma@ttu.edu).

strategy of different drivers on mixed traffic are analyzed considering driving behaviors heterogeneity.

The remainder of this paper is organized as follows. In Section II, the inverse reinforcement learning-based driver behavior model is developed. In Section III, the personalized ACC in mixed traffic is derived. In Section IV, the impacts of the proposed personalized ACC strategy on mixed traffic are investigated through numerical simulations in real-world driving cycles. At last, concluding remarks are made in Section V.

## II. Driver Behavior Modeling

In this section, the feature-based driver behavior model is proposed to model the human driver's behaviors. The goal is to acquire a unique cost function from the driver's demonstrated driving data that best explains the driver's preferred driving behaviors. This unique cost function can be used to model driver's behaviors and to plan longitudinal vehicle operations in real traffic scenarios.

### A. Trajectory Representation

The vehicle trajectory model, represented as the longitudinal position of the human-driven vehicle, is defined as a one-dimensional quantic polynomial in a time interval $[t_k, t_k + T_h]$, $k = 0, 1, \ldots N-1$, for a trajectory with $N$ segments and $T_h$ defines the length of each trajectory segment. The longitudinal position of the human-driven vehicle for each trajectory segment is defined as in (1)

$$r(t) = c_0 t^5 + c_1 t^4 + c_2 t^3 + c_3 t^2 + c_4 t + c_5 \tag{1}$$

where $t \in [t_k, t_k + T_h]$ and $c_{0-5}$ are the coefficients that provide six degrees of freedom for each demonstrated trajectory segment. The longitudinal velocity and acceleration of the human-driven vehicle can be expressed as $\dot{r}(t)$ and $\ddot{r}(t)$ respectively.

### B. Inverse Reinforcement Learning (IRL)

In this study, the IRL approach is used to learn human driver behavior models from the demonstrations. By using the set of $N$ observed trajectory segments, the goal is to learn a driver behavior model that describes the observations with $N$ most likely trajectory segments representing the complete trajectory. Therefore, a set of features is used to capture the relevant characteristics of an individual's driving style. Each feature $f_n$ is a function that represents a real scalar value of the corresponding driver behavior characteristics within the trajectory. According to the IRL approach, the cost function can be defined as a linear combination of the features and their corresponding weights as shown in (2)

$$J = \mathbf{W}^T \mathbf{f}(r) \tag{2}$$

where $J$ is the cost function; $\mathbf{W} = (W_1, W_2, \cdots, W_n)^T$ is the weight vector; $\mathbf{f}(r) = (f_1, f_2, \cdots, f_n)^T$ is the feature vector; $n$ represents the number of defined features.

The goal is to find the weight vector that the expected feature values match the observed feature values as shown in (3)

$$\mathbf{f}^e = \tilde{\mathbf{f}} \tag{3}$$

where $\mathbf{f}^e$ is the expected feature values and $\tilde{\mathbf{f}}$ is average observed feature values of the demonstrated trajectory as shown in (4)

$$\tilde{\mathbf{f}} = \frac{1}{N} \sum_{i=1}^{N} \mathbf{f}(\tilde{r}_i) \tag{4}$$

where $N$ represents the number of trajectory segments in the trajectory set.

According to the principle of Maximum Entropy [13], the probabilistic model that yields a probability distribution over trajectory $p(r | \mathbf{W})$ is proportional to the negative exponential of the costs obtained along the trajectory as shown in (5)

$$p(r | \mathbf{W}) = \exp(-\mathbf{W}^T \mathbf{f}(r)) \tag{5}$$

where agents are exponentially more likely to select trajectories with lower cost $\mathbf{W}^T \mathbf{f}(r)$. It is usually not possible to derive $\mathbf{W}$ analytically, but the gradient of the optimization problem with respect to $\mathbf{W}$ can be derived, as stated in [10]. The gradient can be computed as the difference between the expected and observed feature values as shown in (6)

$$\nabla \mathbf{f} = \mathbf{f}^e - \tilde{\mathbf{f}} \tag{6}$$

The expected feature values can be computed as the feature values of the most likely trajectory by maximizing $p(r | \mathbf{W})$ as shown in (7)

$$\mathbf{f}^e \approx \mathbf{f}(\arg\max p(r | \mathbf{W})) \tag{7}$$

The feature weight vector can be updated by the normalized gradient descent (NGD) method [11] as shown in (8)

$$\mathbf{W} \leftarrow \mathbf{W} + \eta \frac{\nabla \mathbf{f}}{\|\nabla \mathbf{f}\|} \tag{8}$$

where $\eta$ is the learning rate.

### C. Feature Construction

In this section, the features which are shown capable of capturing primary longitudinal driving behaviors [11] are used to represent relevant characteristics of driving behaviors:

**Acceleration:** The integration of the acceleration over each trajectory segment is considered as a feature to capture acceleration and deceleration motions.

$$f_a(t) = \int_t^{t+T_h} \|\ddot{r}(t)\|^2 dt \tag{9}$$

**Desired Speed:** The integration of the deviation from the desired speed is used to identify the driver's preference for achieving and cruising at the driver's particular speed. The desired speed $v_d$ is set to the observed maximum speed of the driver within the trajectory segment.

$$f_{ds}(t) = \int_t^{t+T_h} \|v_d - \dot{r}(t)\|^2 dt \tag{10}$$

**Relative Speed:** The integration of the relative speed is used to describe the driver's preference for following the preceding vehicle speed $v_p$ and maintaining a constant gap distance.

$$f_{rs}(t) = \int_t^{t+T_h} \|v_p(t) - \dot{r}(t)\|^2 dt \tag{11}$$

**Relative Distance:** The integration of the gap distance variation from the desired value $d_D$ is used to identify the driver's preference for maintaining the gap distance $d(t)$ where $d_s$ is the minimum safety clearance and $\tau$ is the time headway. $\tau$ is defined as the observed minimum time headway from the driver's demonstrated trajectory segments.

$$d_D = \dot{r}(t)\tau + d_s \quad (12)$$

$$f_{rd}(t) = \int_t^{t+T_h} \|d(t) - d_D\|^2 \, dt \quad (13)$$

### D. Driver Behavior Learning Algorithm

In this section, the details of the driver behavior learning algorithm will be illustrated. The trajectory set is divided into the trajectory segments $(r_1, r_2, \cdots r_N)$ and the following steps of the algorithm are applied for the driver behavior learning process as shown in Algorithm 1.

---
**Algorithm 1: Driver Behavior Learning Algorithm**

**Input:** $(r_1, r_2, \cdots r_N)$
**Output:** $\mathbf{W}$, $(r_1^*, r_2^*, \cdots r_N^*)$
1: $\mathbf{W} \leftarrow$ all-ones vector
2: $\tilde{\mathbf{f}} = \frac{1}{N}\sum_{i=1}^{N} \mathbf{f}(\tilde{r}_i)$
3: **while** $\mathbf{W}$ not converged **do**
4:      **for** all $r_i \in (r_1, r_2, \cdots r_N)$ **do**
5:          $(c_5, c_4, c_3) \leftarrow$ (position, velocity, acceleration)
6:          at the initial state of the $r_i$
7:          Optimize $(c_2, c_1, c_0)$ with respect to $\mathbf{W}^T \mathbf{f}$
8:      **end for**
9:      $\mathbf{f}^e = \frac{1}{N}\sum_{i=1}^{N} \mathbf{f}(r_i^*)$
10:     $\nabla \mathbf{f} = \mathbf{f}^e - \tilde{\mathbf{f}}$
11:     $\mathbf{W} \leftarrow \mathbf{W} + \eta \frac{\nabla \mathbf{f}}{\|\nabla \mathbf{f}\|}$
12: **end while**

---

## III. PERSONALIZED ADAPTIVE CRUISE CONTROL (PACC) IN MIXED TRAFFIC

In the previous section, we learned the cost function that best represents the driver's preferences using the demonstrated driving data. Next, we will design a personalized ACC (PACC) for the semi-autonomous vehicle (SAV) to plan its longitudinal operations in the car-following scenario. The proposed design minimizes the learned unique cost function by utilizing the preview information of the preceding vehicle (PV) where the SAV is assumed to able to access the perfect knowledge of the short-horizon speed preview information of the PV through vehicle connectivity. Besides, a homogenous human-driven fleet is described as the following traffic for the SAV to assess the impacts of the proposed design on mixed traffic.

The concept of the proposed design in mixed traffic is shown in Fig. 1.

### A. Inter-vehicle Dynamics Model

In this section, a simple inter-vehicle dynamics model is implemented for the longitudinal speed planning purpose of the SAV [14]. The discretized inter-vehicle dynamics model with sample time $T_s$ is used for state updating at time step $j+1$ as in (14)

$$\begin{aligned}\Delta x(j+1) &= (V_{PV}(j) - V_0(j))T_s + \Delta x(j) \\ V_0(j+1) &= V_0(j) + a_0(j)T_s\end{aligned} \quad (14)$$

where $\Delta x$ is the gap distance between the SAV and the PV, $V_0$ and $V_{PV}$ are the longitudinal velocity of the SAV and the PV, respectively, $a_0$ is the longitudinal acceleration of the SAV.

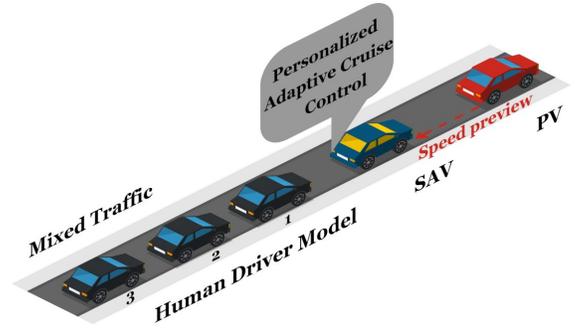

Fig. 1 Schematic of the personalized adaptive cruise control in mixed traffic.

### B. Nonlinear Model Predictive Control (NMPC) Implementation

In this section, the nonlinear model predictive control (NMPC) algorithm is introduced to solve the PACC due to the nonlinearity introduced by the cost function. The basic concept of NMPC is to utilize the numerical optimization techniques and the predictive model to derive a control input sequence that minimizes the specified cost function over a defined time horizon subject to the system constraints [15].

In this study, the design objective of the NMPC is to compute the optimal acceleration of the SAV within the control horizon $T_C$ by minimizing the cost function $J$, which is an individual's unique cost function over the prediction time horizon $T_P$ subject to the system constraint. The cost function and constraint are defined as:

$$J = \sum_{j}^{j+T_P/T_s} \mathbf{W}^T \mathbf{f} \quad (15)$$

$$\begin{aligned}&\text{subject to}: \\ &\Delta x_{\min} \leq \Delta x(j)\end{aligned} \quad (16)$$

where $\mathbf{f} = (f_a, f_{ds}, f_{rs}, f_{rd})^T$ and $\Delta x_{\min}$ is the minimum gap distance that guarantees the safety clearance during the car-following scenario and it is defined as 5 m. At each time step, the proposed NMPC design solves the nonlinear program with the sequential quadratic programming (SQP) algorithm [16] and only the first value of the control vector that includes SAV's acceleration corresponding to the actual time step is applied to the system. In the NMPC design, prediction time $T_P$ and control horizon $T_C$ are set to 3 s with sample time $T_S$ 1 s.

*C. Intelligent Driver Model*

In this study, the Intelligent Driver Model (IDM) [17] is used to describe car-following behaviors of the human-driven vehicles when following the SAV in mixed traffic as shown in Fig. 1. The model is briefly formulated as follows:

$$\dot{v}_i = a_i \left(1 - \left(\frac{v_i}{v_{s_i}}\right)^\delta - \left(\frac{s^*(v_i, \Delta v_i)}{s_i}\right)^2\right) \quad (17)$$

$$s^*(v_i, \Delta v_i) = s_0 + Tv_i + \frac{v_i \Delta v_i}{2\sqrt{ab}} \quad (18)$$

where $v$ is vehicle velocity; subscript $i$ represents the $i$th vehicle in the fleet; $a$ is the maximum acceleration; $b$ is the maximum deceleration; $\delta$ is the acceleration component; $v_s$ is the cruising speed; $\Delta v$ is the speed difference to the preceding vehicle; $s^*$ is the desired inter-vehicle distance; $s_0$ is the jam distance; $T$ is the desired time headway. In this study, we consider a homogenous human-driven fleet that consists of three human-driven vehicles. The IDM model parameters are selected from the realistic range of driving preference parameters [18]. The maximum acceleration $a$ and deceleration $b$ are set to 2 m/s² and 3 m/s² respectively; the desired time headway $T$ is set to 1 s; the jam distance $s_0$ is set to 2 m; the cruising speed $v_s$ is set to the maximum speed of the SAV along the trip; the acceleration component $\delta$ is set to 4.

## IV. RESULTS AND DISCUSSIONS

*A. Data Set and Learning Model Configuration*

In this paper, we use two different driver's driving data from the Human-AV Interaction Trajectory data set [19] to train and test the driver behavior model. The dataset consists of two leader-follower trajectories for each driver where the leader is either a human-driven vehicle or an automated vehicle, and the follower is a human-driven vehicle. The data is collected from the following human-driven vehicles at 10 Hz through an in-field experiment. In this study, we only aim to model the human-driven vehicle under the car-following scenario where the leader vehicle is another human-driven vehicle. In this case, we only used the leader-follower trajectories from the dataset where the leader vehicle is human-driven. Therefore, a total of 8 leader-follower trajectories (4 trajectories for each driver) are used to develop the driver behavior model.

To evaluate the learning performance of the model and its ability to generalize over two different drivers in various speed profiles, we randomly first select three trajectories for each driver for the training, and the remaining one trajectory is used for the testing. For trajectory optimization with respect to spline parameters as in step 7 of the driver behavior learning algorithm, the nonlinear unconstrained optimization method in MATLAB/Optimization Toolbox is used. In the weight vector update, the learning rate $\eta$ is set to 0.2 at the initial and then drops by half for every five epochs. The length of each trajectory segment $T_h$ is set to 3 seconds, and the safe distance between the vehicles $d_s$ is set to 5 m.

*B. Assessment of Driver Behavior Learning Model*

In this section, the performance of the designed driver behavior learning model will be investigated through two different drivers in the car-following scenarios. The learning method aims at finding weight vectors **W** that represent the car-following strategies of each driver.

Fig. 2 shows the actual and predicted speed profiles of both drivers in the testing. It is shown that the driver model generates accurate speed profiles with the learned cost function against the ground truth. Table I shows the RMSE values between the observed and the predicted trajectories for testing using the learned cost functions of each driver. We find that the proposed method can replicate the observed trajectories for the drivers with small prediction errors. Since the expected features are computed as the feature values of the most likely trajectories during the learning process, it is acceptable to observe minor discrepancies between the demonstrated and predicted trajectories during the testing. The results indicate that driving characteristics and behaviors can be consistently captured and replicated with the proposed modeling approach across different drivers.

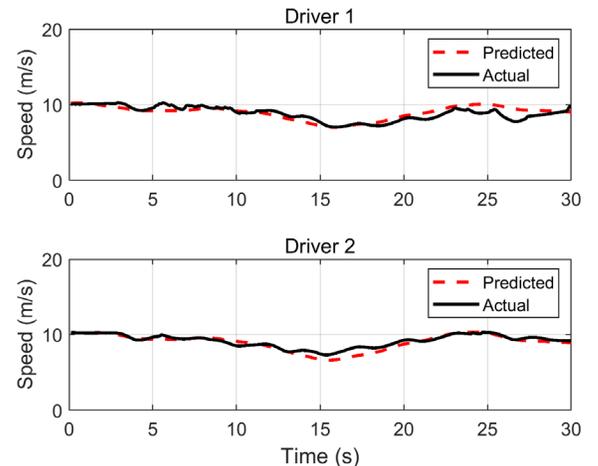

Fig. 2  Speed profiles of Driver 1 and Driver 2 for the testing.

Table I:  RMSE values.

| Driver | Position (m) | Speed (m/s) |
|---|---|---|
| 1 | 0.935 | 0.716 |
| 2 | 0.692 | 0.447 |

The learned feature weights of each driver, which represents an individual's driving preferences, are listed in Table II. The learned feature weights significantly demonstrate that human drivers have different driving preferences when they have free control of vehicle motions, even under the same driving conditions.

Table II: Learned feature weights.

| Driver | Acceleration | Desired Speed | Relative Speed | Relative Distance |
|---|---|---|---|---|
| 1 | 0.989 | 0.064 | 0.962 | 0.112 |
| 2 | 0.970 | 0.055 | 0.867 | 0.132 |

### C. Personalized Adaptive Cruise Control (PACC) Impacts on Mixed Traffic

By utilizing the previously developed models, we want to evaluate the impacts of the SAV with different drivers on mixed traffic. Therefore, we designed two different scenarios for comparative analysis as we introduce scenarios A and B. In both scenarios, the SAV is operated with different drivers and follows the PV that operates according to the predefined speed profile, and a homogenous human-driven fleet that is described by the IDM model follows the SAV in mixed traffic. In Scenario A and B, the SAVs that are operated by two different drivers are denoted as SAV1 and SAV2, respectively. For the sake of conciseness, the human-driven vehicles in the fleet are denoted as "HV1", "HV2" and "HV3" in both scenarios.

To analyze the impacts of the SAV1 and SAV2 on mixed traffic, we derived average gap distance and time headway, and total trip fuel consumption of mixed traffic for each scenario. The average gap distance and time headway of traffic is derived as shown in (19) and (20)

$$\Delta x_M = \frac{1}{4}\sum_{i=1}^{4}\Delta x_{a_i} \quad (19)$$

$$T_M = \frac{1}{4}\sum_{i=1}^{4} T_{a_i} \quad (20)$$

where $\Delta x_M$ is the average gap distance of traffic; subscript $i$ represents the $i$th pair of consecutive vehicles in the traffic; $\Delta x_a$ is the average gap distance between the consecutive vehicles along the trip; $T_M$ is the average time headway of traffic; $T_a$ is the average time headway between the consecutive vehicles along the trip. The instantaneous vehicle fuel consumption is calculated by Virginia Tech Comprehensive Power-based Fuel Consumption Model (VT-CPFM), a power-based fuel consumption model as formulated in [20]. The total trip fuel consumption of traffic is derived in (21)

$$F_C = \sum_{k=1}^{5}\int_{t_0}^{t_f} \dot{m}_{f_k} dt \quad (21)$$

where $F_C$ is the total fuel consumption of traffic ($L$) during the trip starting at $t_0$ and ending at $t_f$; $\dot{m}_f$ is the instantaneous fuel rate ($L/s$); subscript $k$ represents the $k$th vehicle in traffic.

In this study, we conducted an experiment to get real driving data from traffic during daily commuting. The speed profile of the PV is defined with this daily commute driving cycle. We use this cycle to represent a realistic traffic scenario where the drivability of the PV is guaranteed. The speed profile can be found in Fig. 3 (PV).

Fig. 3 and Fig. 4 show the speed and gap distance profiles of the traffic participants in Scenario A and B, respectively. We find that the SAV2 follows the PV in a much more aggressive manner compared with the SAV1 by comparing the gap distance profiles in Fig. 4 where the gap distance of the SAV2 is found to be much lower than the SAV1. According to the results in Table III, significant differences are observed at the average gap distance and the time headway of traffic when comparing the traffic flow of Scenario A and B. It is found that the SAV2 can significantly reduce average gap distance and time headway of traffic with minor sacrifice in fuel economy. Now, we recall the fact that human drivers may have different driving behaviors when operating the vehicles in traffic as we discussed in the previous section, SAV1 and SAV2 show that such discrepancies will influence the traffic flow of mixed traffic differently. SAV2's more aggressive driving behaviors compared to SAV1 can afford to improve traffic flow by reducing the average gap distance and time headway of mixed traffic despite the reasonable increase in total trip fuel consumption.

Table III: Traffic flow comparison in Scenario A and B.

| | Average Gap Distance | Average Time Headway | Total Trip Fuel Consumption |
|---|---|---|---|
| Scenario A | 25.76 m | 1.68 s | 6.46 L |
| Scenario B | 21.37 m | 1.40 s | 6.57 L |
| **Difference** | **17.03 %** | **16.48 %** | **-1.79 %** |

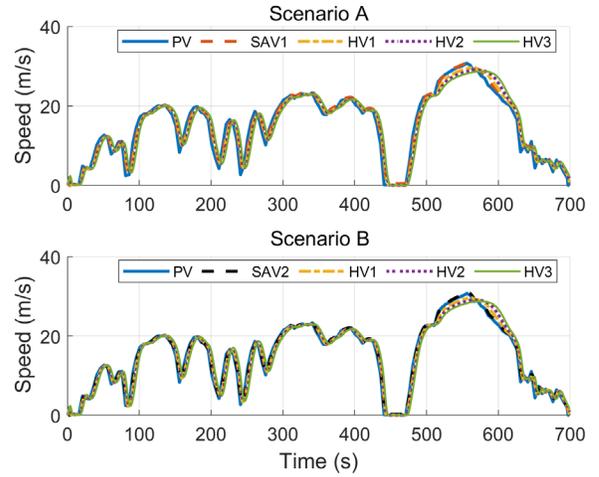

Fig. 3 Speed profiles of the traffic participants in Scenario A and B.

To analyze the impacts of the proposed PACC design with different drivers on mixed traffic, the different real-world driving cycles [21] are tested as well. We use the US06 and FTP-75 driving cycles which are designed by the US Environmental Protection Agency (EPA) for vehicle fuel economy and emission tests. According to the results in Table IV, SAV2 provides a 15-17% reduction in average gap distance and a roughly 16% in average time headway of traffic with a 2-5% sacrifice in fuel economy when comparing the traffic flow of Scenario A and B. These findings suggest that the proposed PACC design is suitable to adaptively control the vehicle with similar observed driving styles from human drivers in different traffic scenarios. Additionally, the extent of such impacts of PACC on mixed traffic varies among the tested drivers due to their unique driving characteristics.

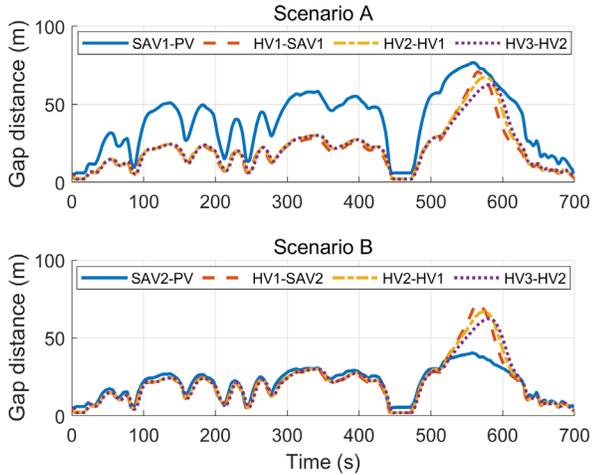

Fig. 4  Gap distance comparison in Scenario A and B.

Table IV:  Traffic flow difference in Scenario A and B for real-world driving cycles.

| Driving Cycle | Average Gap Distance | Average Time Headway | Total Trip Fuel Consumption |
|---|---|---|---|
| US06 | 17.06 % | 16.43 % | -5.59 % |
| FTP-75 | 15.66 % | 16.40 % | -2.32 % |

## V.  Conclusion and Future Work

In this study, we designed a personalized adaptive cruise control strategy for semi-autonomous vehicles to learn and replicate driver's preferred driving behaviors in real traffic scenarios. Moreover, the impacts of the personalized adaptive cruise control on mixed traffic are qualitatively investigated through simulation cases considering individual drivers' behavior variations, which are modeled from the demonstrated driving data of two different drivers using the inverse reinforcement learning approach. Results show that the proposed personalized adaptive cruise control design is capable of characterizing and replicating the car-following strategies of the drivers effectively in a variety of traffic situations. Furthermore, the statistically different impacts of the proposed personalized adaptive cruise control strategy by different drivers on mixed traffic are discovered owing to drivers' intrinsic driving preferences. As a further extension of the work in the future, we will investigate the impacts of the personalized adaptive cruise control system with different levels of penetration rates on mixed traffic to have a more comprehensive statistical investigation.

## References

[1] S. Choi, F. Thalmayr, D. Wee and F. Weig, 2018, "Advanced driver-assistance systems: Challenges and opportunities ahead," Retrieved from https://www.mckinsey.com/industries/semiconductors/our-insights//advanced-driver-assistance-systems-challenges-and-opportunities-ahead

[2] H. Winner, S. Witte, W. Uhler, and B. Lichtenberg, "Adaptive Cruise Control System: Aspects and Development Trends," SAE Paper 961010, 1996.

[3] P. Fancher, Z. Bareket, and R. Ervin, "Human-centered design of an ACC-with-braking and forward-crash-warning system," *Veh. Syst. Dyn.*, vol. 36, no. 2–3, pp. 203–223, 2001.

[4] A. Vahidi and A. Eskandarian, "Research advances in intelligent collision avoidance and adaptive cruise control," in *IEEE Transactions on Intelligent Transportation Systems*, vol. 4, no. 3, pp. 143-153, Sept. 2003.

[5] M. Hasenjäger, M. Heckmann and H. Wersing, "A Survey of Personalization for Advanced Driver Assistance Systems," in *IEEE Transactions on Intelligent Vehicles*, vol. 5, no. 2, pp. 335-344, June 2020.

[6] J. Wang, L. Zhang, D. Zhang and K. Li, "An Adaptive Longitudinal Driving Assistance System Based on Driver Characteristics," in *IEEE Transactions on Intelligent Transportation Systems*, vol. 14, no. 1, pp. 1-12, March 2013.

[7] X. Chen, Y. Zhai, C. Lu, J. Gong and G. Wang, "A learning model for personalized adaptive cruise control," *2017 IEEE Intelligent Vehicles Symposium (IV)*, Los Angeles, CA, 2017, pp. 379-384.

[8] A. P. Bolduc, L. Guo and Y. Jia, "Multimodel Approach to Personalized Autonomous Adaptive Cruise Control," in *IEEE Transactions on Intelligent Vehicles*, vol. 4, no. 2, pp. 321-330, June 2019.

[9] D. Nava, G. Panzani, P. Zampieri and S. M. Savaresi, "A personalized Adaptive Cruise Control driving style characterization based on a learning approach," *2019 IEEE Intelligent Transportation Systems Conference (ITSC)*, Auckland, New Zealand, 2019, pp. 2901-2906.

[10] M. Kuderer, S. Gulati, and W. Burgard, "Learning driving styles for autonomous vehicles from demonstration". In 2015 IEEE International Conference on Robotics and Automation (ICRA), pp. 2641–2646.

[11] M.F. Ozkan and Y. Ma, "Inverse Reinforcement Learning Based Driver Behavior Analysis and Fuel Economy Assessment." *Proceedings of the ASME 2020 Dynamic Systems and Control Conference*. Virtual, Online. October 5–7, 2020. V001T02A003.

[12] Q. Zou, H. Li, and R. Zhang, "Inverse reinforcement learning via neural network in driver behavior modeling". In 2018 IEEE Intelligent Vehicles Symposium (IV), pp. 1245–1250, 2018.

[13] B. D. Ziebart, A. Maas, J. A. Bagnell, and A. K. Dey, "Maximum entropy inverse reinforcement learning". In Proc. AAAI, pp. 1433–1438, 2008.

[14] M. F. Ozkan and Y. Ma, "A Predictive Control Design with Speed Previewing Information for Vehicle Fuel Efficiency Improvement," *2020 American Control Conference (ACC)*, Denver, CO, USA, 2020, pp. 2312-2317.

[15] M. F. Ozkan and Y. Ma, "Eco-Driving of Connected and Automated Vehicle with Preceding Driver Behavior Prediction," ASME. *J. Dyn. Sys., Meas., Control*, Sept. 3, 2020.

[16] J. Nocedal and S. J. Wright. *Numerical Optimization*, Second Edition. Springer Series in Operations Research, Springer Verlag, 2006.

[17] M. Treiber, A. Hennecke, and D. Helbing, "Congested traffic states in empirical observations and microscopic simulations," *Phys. Rev. E*, vol. 62, no. 2, pp. 1805–1824, Aug. 2000.

[18] M. Treiber and A. Kesting, "Elementary Car-Following Models," in *Traffic Flow Dynamics*, Springer, Berlin, Heidelberg, 2013, pp. 157–180.

[19] Y. Rahmati, M. K. Hosseini, A. Talebpour, B. Swain and C. Nelson, "Influence of Autonomous Vehicles on Car-Following Behavior of Human Drivers". Transportation Research Record, 2673(12), 367–379, 2019.

[20] S. Park, H. Rakha, K. Ahn and K. Moran, "Virginia Tech Comprehensive Power-Based Fuel Consumption Model (VT-CPFM): Model Validation and Calibration Considerations," International Journal of Transportation Science and Technology, vol. 2, no. 4, pp. 317–336, 2013.

[21] U. S. E. P. A. (EPA), "Dynamometer drive schedules," https://www.epa.gov/vehicle-and-fuel-emissions-testing/ dynamometer-drive-schedules.